\documentclass[twocolumn,aps,prb,graphicx,showpacs]{revtex4}
\usepackage{amsmath}
\usepackage{amsfonts}
\usepackage{amssymb}
\usepackage{graphicx}
\begin{document}
\title{Theory of Low-Temperature Hall Effect in Electron-Doped Cuprates}
\author{Jie Lin and A. J. Millis}
\affiliation{Department of Physics, Columbia University,
538 West 120th Street, New York, NY 10027}

\begin{abstract}
A mean field calculation of the $T\rightarrow 0$ limit
of the  Hall conductance of electron-doped cuprates
such as $Pr_{2-x}Ce_xCuO_{4+\delta}$ is presented.
The data are found to be qualitatively consistent
with the reconstruction of the Fermi
surface expected upon density wave ordering.
The magnitude of the density wave gap
is found to be large.  The Hall resistance
exhibits a nonanalyticity at the
quantum critical point for density wave ordering,
but the amplitude of the anomaly
is found to be unobservably small.
The quantum critical contribution
to $R_H(B)$ is determined.
Quantitative discrepancies between calculation and data remain,
suggesting that the experimental doping is not identical to the
$Ce$ concentration $x$.
\end{abstract}
\maketitle

\section{Introduction}
The unusual properties of high temperature cuprate superconductors
continue to challenge our
understanding of the physics of electrons in metals.
The related ideas of  ``quantum criticality'' \cite{Sachdev03}
and density wave (or ``stripe'') order and
fluctuations \cite{Monthoux91,Kivelson03} have been much debated,
but in the {\em hole-doped} cuprates
the discussion has been inconclusive.
Further, while it has been proposed \cite{Sachdev03,Varma99} that many
of the anomalous properties of the materials could be explained by a
quantum critical point at or near
``optimal doping'', there is no unambiguous evidence of
long ranged order setting in near optimal
doping in the hole-doped cuprates.

The {\em electron-doped} cuprates offer a new perspective on these issues.
In these materials commensurate
$(\pi,\pi)$ spin density wave order has been detected by muon spin
rotation \cite{Luke90} and neutron scattering
measurements. \cite{Matsuda92,Mang04}  The order exists over a
wide range of dopings and vanishes
at a critical doping $x_c$
near the ``optimal'' doping $x=0.16$.  Thus, in this material an
unambiguous quantum critical
point exists, separating an apparently disordered phase from a phase
with a well-defined
($(\pi,\pi)$ commensurate magnetic) order.
It is therefore of interest to examine the effect of the order
and criticality on material properties.
In this paper we consider the low temperature Hall resistance, which
is sensitive to the rearrangment of the Fermi surface caused
by the onset of the spin density wave order.
We apply to the electron-doped cuprate case a mean-field analysis presented
previously in the literature, \cite{Norman03,Chakravarty02,Oganesyan04}
and compare the results to experiments. \cite{Dagan04,Onose01}

The rest of this paper is organized as follows.
Sec. \ref{sec:Model} defines the model to be studied.
The formal solution to Boltzmann equation with relaxation time approximation
for 2-dimensional systems is derived in Sec. \ref{sec:Formalism}.
Sec. \ref{sec:Weak Field Limit} gives the numerical results
for $R_H$ in weak magnetic field and compares them to experimental data.
In Sec. \ref{sec:Critical Behavior} we analyze our formal solution
in the critical region, $x\lesssim x_c$.
The sensitivity of our numerical results to variation of parameters
is studied in Sec. \ref{sec:Sensitivity}.
The possibility of explaining the $x>x_c$ experimental results
by calculating the real part of the self energy due to spin fluctuations
is investigated in Sec. \ref{sec:Normal State Region}.
We conclude our calculations and discuss our results
in Sec. \ref{sec:Conclusion and Discussion}.

\section{Model}
\label{sec:Model}

We consider electrons moving in a square lattice of unit lattice constant
with dispersion
\begin{multline}
\varepsilon_p=-2t_1\left(\cos p_x+\cos p_y\right)+4t_2\cos p_x\cos p_y \\
-2t_3\left(\cos2p_x+\cos2p_y\right)
\label{dispersion}
\end{multline}
The ``canonical'' values of the band parameters are
$t_1=0.38$eV, $t_2=0.32t_1$ and $t_3=0.5t_2$.
As will be shown below, our main results are not sensitive to the precise
parameter choices. We are interested in electron dopings, corresponding to
two dimensional carrier densities per unit cell $n=1+x>1$.

\begin{figure}[htbp]
\includegraphics[width=2.5in]{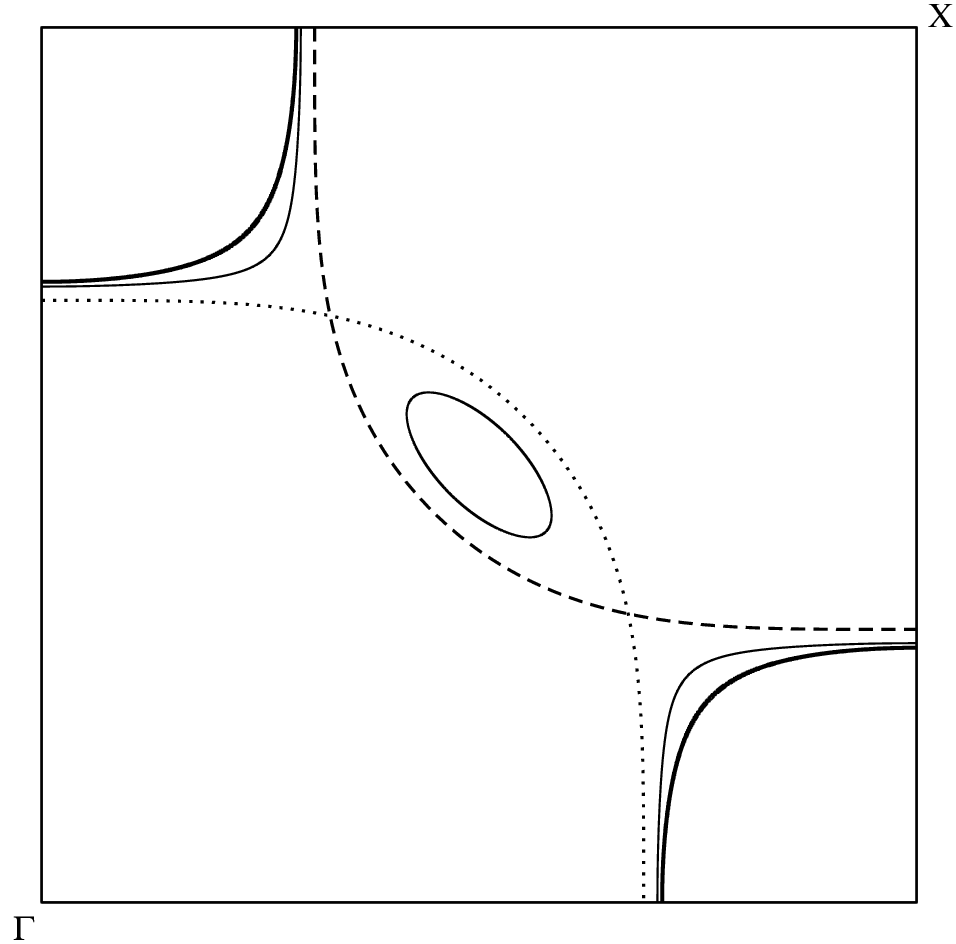}
\caption{\textit{Fermi surfaces calculated from
Eq [\ref{Ep}] for $x=0.15$, band
parameters $t_1=0.38 eV$, $t_2=0.32t_1$, $t_3=0.5t_2$
and backscattering (``gap'') values $\Delta=0$ (dashed line), $\Delta=0.2eV$
(light solid line) and $\Delta=0.4eV$ (heavy solid line).
The dotted line is given by $\varepsilon_{p+Q}=\mu(\Delta=0)$.}}
\label{fig:Fermi surface}
\end{figure}

We assume that for $x$ less than a critical value $x_c$
a commensurate spin density wave (SDW)
order occurs, so the electrons are subject to
a coherent backscattering of amplitude
$\Delta$ and wavevector $\vec{Q}=(\pi,\pi)$ which doubles the unit
cell, changes the dispersion to
\begin{equation}
E_p^{\pm}=\frac{1}{2}\left(\varepsilon_p+\varepsilon_{p+Q}
\pm\sqrt{\left(\varepsilon_p-\varepsilon_{p+Q}\right)^2+4\Delta^2}\right)
\label{Ep}
\end{equation}
and reconstructs the Fermi surface. If $\Delta=0$ then
(at the relevant dopings $x$)
the Fermi surface consists of one large hole surface
centered at the $X$ point $(\pi.\pi)$
(dashed line in Fig \ref{fig:Fermi surface}).
If $\Delta \neq 0$ then the
Fermi surface is reconstructed. The details depend on the band filling
and the magnitude of $\Delta$.
If the chemical potential $\mu$ is such that
$\Delta<\Delta^*(\mu)=\varepsilon_{(\pi/2,\pi/2)}-\mu=4t_3-\mu$ then
both bands cross the Fermi level
and the Fermi surface involves two symmetry-inequivalent
hole pockets centered at $(\pi/2,\pm\pi/2)$
and one electron pocket centered at $(0,\pi)$.
One of the hole pockets and two portions of the electron pocket are
shown as the light solid lines in Fig \ref{fig:Fermi surface}.
However, if $\Delta>\Delta^*(\mu)$,
the lower band is completely filled
and only the electron pocket remains (heavy solid line
in Fig \ref{fig:Fermi surface}). For the parameters
used in Fig \ref{fig:Fermi surface}, $\Delta^*\approx 0.26$eV.

As $x$ is decreased below $x_c$
we assume that the gap magnitude increases.
However, the band filling and hence the
chemical potential decreases. Therefore,
depending on the relative growth (with decreasing $x$)
of $\mu$ and $\Delta$
either the hole pockets vanish at a critical $x=x_1$,
or it persists to $x=0$.
The comparison to data given below suggest that
the former circumstance obtains in the actual materials,
and that indeed the gap opens so rapidly that
$x_1$ is of order $0.12$,
not much less than $x_c\approx 0.16$.

\section{Calculation of Hall Resistance: Formalism}
\label{sec:Formalism}

We require the $T\rightarrow 0$ transport coefficients.
In the $T \rightarrow 0$ limit
the only scattering is due to impurities, and we assume a
Boltzmann equation description is adequate, so that the physics
is described by a momentum space
distribution function $g(p)=f(p)+h_p$ with
$f(p)$ the Fermi-Dirac distribution.
We also adopt the relaxation time
approximation, linearize in the applied electric field ${\vec E}$
and take the $T \rightarrow 0$ limit so that for each band $b$,
the Boltzmann equation becomes ($e=|e|$):
\begin{equation}
-e{\vec E}\cdot {\vec v}^b\delta(E^b_p-\mu)
=\left(\frac{\hbar}{\tau_p}+
\frac{e}{c}{\vec B}\cdot
({\vec v}^b_p \times
\frac{\partial}{\partial {\vec p}})\right)h^b_p
\label{be}
\end{equation}
We are interested in two dimensional electrons. We
take $B \parallel {\hat z}$ and note that
${\vec B}\cdot({\vec v}^b_p \times \frac{\partial}{\partial {\vec p}})$
corresponds to a derivative
along the Fermi contour.
Defining $s$ to be the coordinate along the Fermi arc,
noting that the structure of the equation implies
\begin{equation}
h^b_p=\delta(E^b_p-\mu)h^b(s)
\end{equation}
and temporarily dropping, for ease of writing,
the band superscript b
allow us to rewrite Eq [\ref{be}] as
\begin{equation}
\left[1+a(s)\partial_s\right]h(s)=E(s)
\label{be2}
\end{equation}
with ($l=v\tau$ is the mean free path,
and $\Phi_0=hc/2e$ is the superconducting flux quantum):
\begin{align}
a(s)&=\frac{e \tau(s) B |v(s)|}{\hbar c} =\frac{\pi l(s) B}{\Phi_0}
\label{adef} \\
E(s)&=-\frac{e\tau(s){\vec E}\cdot {\vec v}}{\hbar}
\label{Edef}
\end{align}

Eq [\ref{be2}] may be formally solved \cite{Chambers56}
in terms of the Green function $K(s,s')$ satisfying
\begin{equation}
\left[1/a(s)+\partial_s\right]K(s,s')=\delta(s-s')
\label{Kdef}
\end{equation}
with boundary condition
\begin{equation}
K(s\rightarrow s'^+,s')-K(s\rightarrow s'^-,s')=1
\end{equation}
as
\begin{equation}
h(s)=\oint ds'K(s,s')E(s')/a(s^{\prime})
\label{hsoln}
\end{equation}
The currents longitudinal and transverse to ${\vec E}$
may then be constructed
by averaging over the velocity as usual.
Assuming a square lattice, restoring the
factors of velocity and the band label (if necessary),
and symmetrizing or antisymmetrizing for $\sigma_{xx,xy}$,
we find
\begin{align}
\sigma_{xx}&=\sigma_{Q}\oint \frac{dsds'}{p_B^2}
K(s,s')\cos[\phi(s)-\phi(s')] \label{sigxx}\\
\sigma_{xy}&=\sigma_{Q}\oint \frac{dsds'}{p_B^2}
K(s,s') \sin[\phi(s)-\phi(s')] \label{sigxy}
\end{align}
 where $\sigma_{Q}\equiv e^2/\hbar$ is conductance quantum,
$p_B\equiv 2\pi/l_B$ is proportional to the inverse of the magnetic length
$l_B\equiv (\Phi_0/\pi B)^{1/2}$,
and $\phi$ is the polar angle of Fermi velocity
on the corresponding Fermi contours.
The formulae are valid as long as magnetic breakdown can be neglected.
A criterion for this is given below.

\begin{figure}[htbp]
\centering
\includegraphics[width=2.5in]{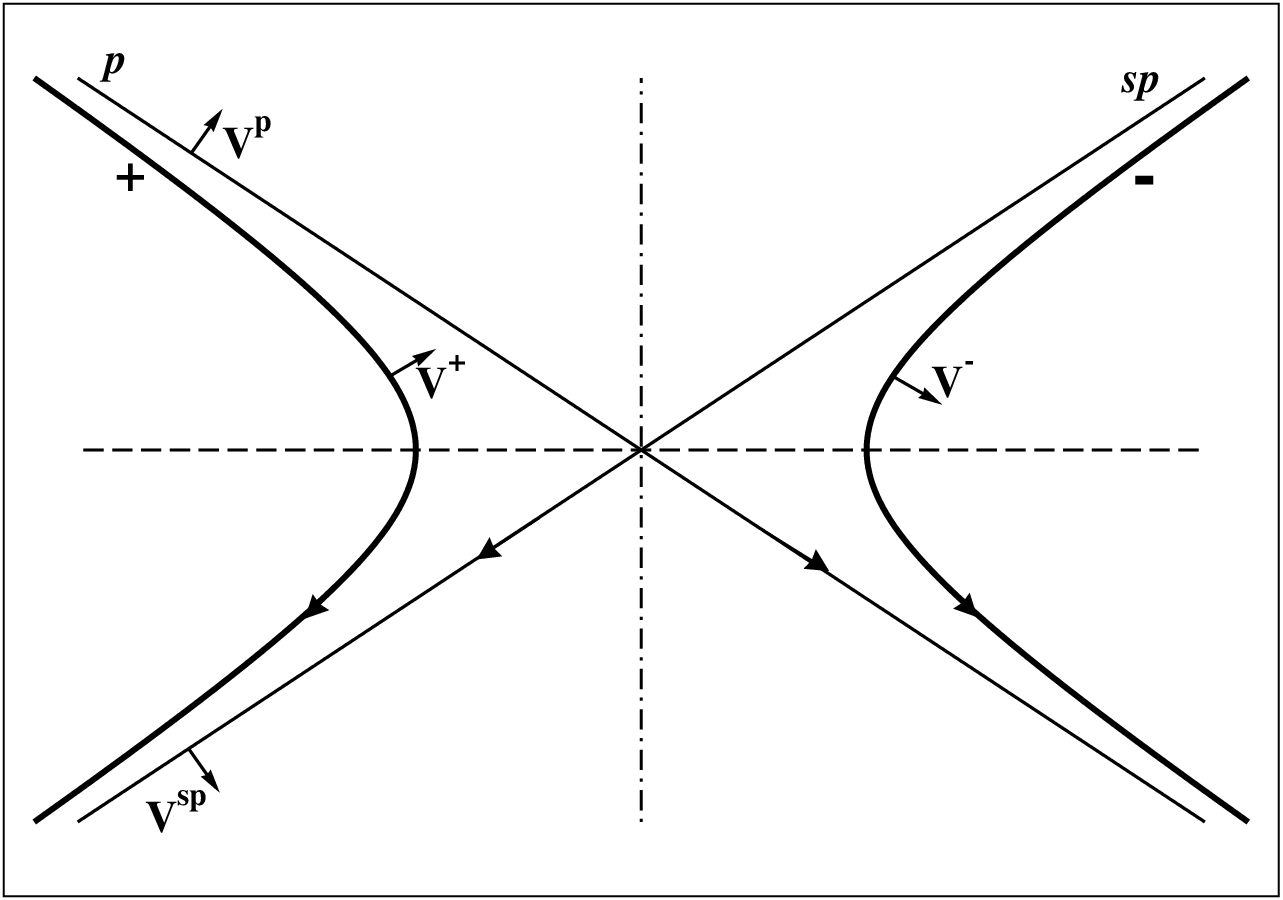}
\caption{\textit{Local view of the Fermi surface
in the vicinity of one of crossing points,
$\varepsilon_p=\varepsilon_{p+Q}=\mu$,
shown in Fig \ref{fig:Fermi surface}.
p, sp are respectively the Fermi lines
for the normal state and the normal state Fermi line shifted by ${\vec Q}$.
The curvature has been neglected.
$+$ denotes the electron pocket of SDW state
and $-$ denotes the hole pocket.
Arrows along Fermi lines show the circulation direction
determined by ${\bf B}\times {\bf v}$ with ${\bf B}\parallel -{\bf z}$.
They also indicate the direction of increase in arclength coordinate,
$s$. $V^p$, $V^{sp}$, $V^+$ and $V^-$ are Fermi velocities
along corresponding Fermi lines.
The dashed line is the diagonal of the first quadrant of Brillouin zone
connecting $(0,\pi)$ and $(\pi,0)$.
The dot-dashed line is perpendicular to the dashed line:
if the normal state energy dispersion is linearized,
and the momentum dependence of velocity is neglected,
the problem is symmetric under reflection through the dot-dashed line.}}
\label{fig:localview}
\end{figure}

The problem has several momentum scales.
Obtaining a non-vanishing Hall coefficient requires
breaking of particle-hole symmetry,
either by a variation of scattering rate,
a Fermi surface curvature, or a variation of velocity.
All of these phenomena are characterized by momentum scales
of the order of the size of the Brillouin zone;
we refer to these scales generically as $p_0$.
We take these to be large, so we may neglect Landau quantization.
A second important scale $p_{\Delta}=\Delta/v_F$ is set by
the SDW backscattering amplitude, $\Delta$,
and vanishes as $\Delta\rightarrow 0$.
A third scale is the range, $a$, of the kernel, $K$ (Eq [\ref{adef}]).
We distinguish weak field ($a<p_{\Delta}$)
and strong field ($a>p_{\Delta}$) regimes.
A fourth significant scale is the inverse of the mean free path, $l^{-1}$.
The results we present are valid for $p_0\gg p_{\Delta},a,l^{-1}$.
In addition, we expect disorder to modify the SDW behavior
significantly for $p_{\Delta}l\sim 1$;
so our results do not apply for $\Delta<v_F/l$.
In this case we may neglect the periodicity and write
\begin{equation}
K(s,s')=\exp\left[-\int_{s'}^s\frac{dx}{a(x)}\right]\Theta(s-s')
\label{Ksoln}
\end{equation}

Finally, we note that all of the formulae are derived on the assumption
that a state evolves along a given Fermi contour
and does not jump to another one,
{\it i.e.}, that magnetic breakdown may be neglected.
By applying the considerations of Ref. \onlinecite{Blount61}
to the model studied here, we find
the condition for absence of magnetic breakdown is,
in order of magnitude,
that the inverse of the magnetic length $p_B=(\pi B/\Phi_0)^{-1/2}2\pi$
be less than the gap momentum scale $p_{\Delta}$.
Our results are only valid if $a,p_{\Delta}>p_B$.
The condition that breakdown occurs at a field higher than
the weak-strong field crossover is $p_{\Delta}l>1$,
which as noted above is a necessary condition
for the validity of the approach.

\section{Weak Field Limit: Evaluation and Comparison to Data}
\label{sec:Weak Field Limit}

In this section we analyse in detail the case
in which the field is weak enough
that $a$ is small relative to all scales in the problem.
In this case the $s'$ integrals in Eqs [\ref{sigxx},\ref{sigxy}]
are dominated by $s' \approx s$ and may be performed, leaving
\begin{align}
\sigma_{xx}&=\sigma_{Q}\frac{1}{4\pi^2}\oint ds  l(s)
\label{sigxxlow}\\
\sigma_{xy}&=\sigma_{Q}\frac{1}{4\pi^2}\oint ds
\left(l(s)/l_B\right)^2\frac{\partial \phi(s)}{\partial s}
\label{sigxylow}
\end{align}
for the longitudinal and Hall conductance per plane.
Eq [\ref{sigxxlow}] is exactly the Drude result
(average of mean free path over the Fermi arc),
while Eq [\ref{sigxylow}] reproduces Ong's result \cite{Ong91}
\begin{equation}
\sigma_{xy} =\sigma_{Q}\frac{B}{\Phi_{0}}\frac{1}{2\pi}
\oint d{\vec l}\times {\vec l}\cdot \hat{{\bf z}}/2
\label{ongformula}
\end{equation}

Further analysis requires information about the scattering rates.
The two commonly made assumptions are:
\begin{itemize}
\item mean free path is constant over the Fermi surface
\item scattering rate is constant over Fermi surface
\end{itemize}
The constant scattering rate assumption may be derived theoretically
by assuming weak (Born approximation) point-like scatterers.
Constant mean free path is more appropriate
for very strong (unitary) scattering.

Ong \cite{Ong91} shows that the constant mean free path
approximation implies that the Hall conductance
for each Fermi surface pocket is a constant, depending only
on the value of the mean free path and the
sign (electron- or hole-like) of the Fermi pocket. Therefore,
in the constant mean free path approximation,
$\sigma_{xy}$ does not change as
$x$ is decreased through $x_c$.
The reconstruction of the Fermi surface means that
one goes from one hole pocket
($\sigma_{xy} \propto l^2$) to two hole pockets
($\sigma_{xy} \propto 2 l^2$) and one electron pocket
($\sigma_{xy} \propto -l^2$) so that the total
$\sigma_{xy} \propto l^2(2-1)=l^2$.
However, at $x=x_1$ the hole pockets
vanish, and in the constant mean free path approximation the Hall
coefficient would jump discontinuously from a value close to the
$R_H \sim 1/(1-x)$ implied by the initial, large, hole-like surface,
to the value $R_H \sim  -1/x$ implied by a single electron-like pocket.
This behavior is in evident disagreement with data,
\cite{Dagan04,Onose01} which
indicates instead a smooth change beginning at $x \approx 0.16$,
which is approximately
the doping at which antiferromagnetism sets in. \cite{Mang04}

\begin{figure}[htbp]
 \centering
 \includegraphics[width=2.5in]{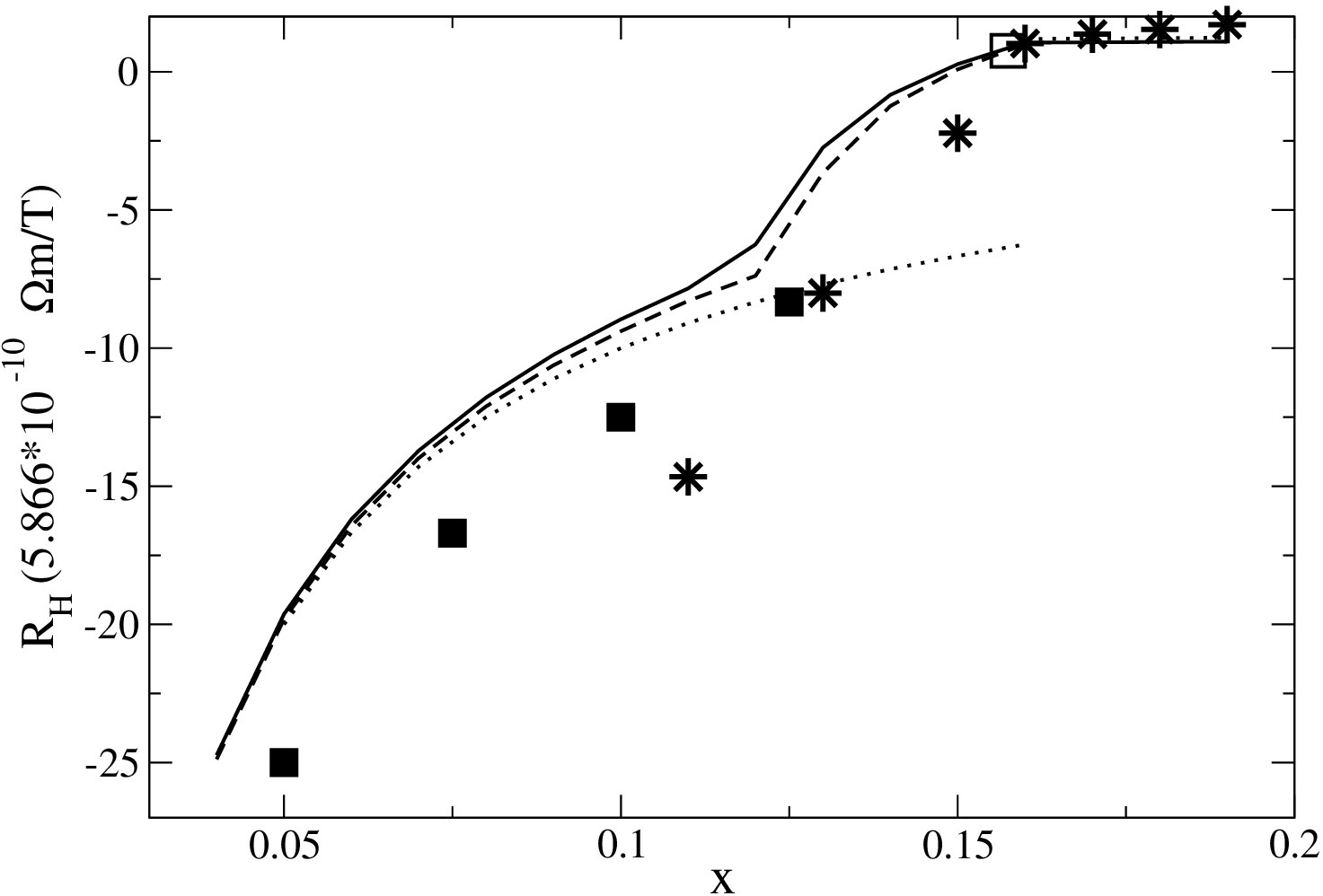}
 \caption{\textit{Solid line: Hall coefficient calculated from Eqs
[\ref{sigxxlow},\ref{ongformula}] for parameters $t_1=0.38$eV,
$t_2=0.32t_1$, $t_2=0.5t_2$ and $\Delta(x)[eV]=0.6\sqrt{1-x/0.16}$.
Long dashed line: Hall coefficient calculated for parameters
$t_1=0.38$eV, $t_2=0.32t_1$, $t_2=0.075$eV and
$\Delta(x)[eV]=0.7\sqrt{1-x/0.16}$.
They are calculated with constant $1/\tau$ approximation.
Dotted line: curve $R_H=-1/x$ for $x<0.16$ and $R_H=1/(1-x)$
for $x>0.16$. Stars: experimental data. \cite{Dagan04}
Solid squares: experimental data. \cite{Onose01}
The open box shows the approximate size of the region
for which the weak field-strong field crossover may be observable
with present accessible fields.}}
\label{fig:hall}
\end{figure}

We therefore turn to the constant $1/\tau$ approximation.
In this case, $\sigma_{xy}$ for each pocket
is weighted by the square of the
Fermi velocity for that pocket and more variation occurs.
The solid line and long dashed line in Fig \ref{fig:hall}
show typical results obtained by evaluating
Eqs [\ref{sigxxlow},\ref{ongformula}]
in the constant scattering rate approximation
and assuming a backscattering amplitude growing
proportional to $\sqrt{x_c-x}$.
One sees that for $x>x_c$ the Hall resistance is
hole-like and depends only very weakly on $x$.
At $x=x_c$ a nonanalyticity occurs, and $R_H$ begins to decrease.
For the parameters chosen, $x_1 \sim 0.12$.
At this doping the hole pockets
vanish, a further change in curvature occurs,
and $R_H$ approximates very closely the $1/x$ behavior
expected from the single remaining electron pocket
(shown as the dotted line).

Fig \ref{fig:hall} also shows, as stars and heavy squares,
the experimental results. \cite{Dagan04,Onose01}
These data are in qualitative agreement with calculation,
but important quantitative differences occur.
At the lowest doping
(nominal carrier concentration $x=0.11$ for stars \cite{Dagan04}
and $x=0.05$ for heavy squares \cite{Onose01})
the measured $R_H$ is substantially below
both the calculated value and the $1/x$ curve.
It seems to us essentially impossible
to obtain a Hall resistance satisfying $|R_H|>|1/x|$;
we therefore suggest that the
nominal doping level, $x$, differs from the actual doping.
The presently available data do not allow us to
locate the point $x_1$ with precision,
but strongly suggest that $x_1\geq0.1$.

\section{Critical Behavior}
\label{sec:Critical Behavior}

In this section, we study in more detail the behavior of Hall
conductance $\sigma_{xy}$ in the vicinity of $x=x_c$. When the
system is tuned through the critical point (which we believe to be
$x_c \approx 0.16$ in the electron-doped cuprates), the coherent
backscattering introduces reconstruction of Fermi surface, changes
its local curvature, and thus affects the Hall conductance. The
weak field limit was previously studied by Bazaliy {\it et al}
\cite{Bazaliy04} in the context of the 3d SDW compound $Cr$. They
obtained a result equivalent to our Eq [\ref{weakfieldsigmaxx}]
and a qualitative result equivalent to the weak field limit of our
Eq [\ref{scalingfunction}]. And the strong field limit was studied
by Fenton and Schofield, \cite{Schofield05} who found a result
equivalent to our Eqs
[\ref{strongfieldresult},\ref{scalingrelaxation}]. The two
dimensional nature of the present problem allows us to carry the
analysis further, obtaining prefactors for $\sigma_{xy}$ in the
weak field limit and a treatment of the weak to strong field
crossover.

To study the vicinity of the critical point, we use Eq
[\ref{sigxy}] to write an equation
for the change in the Hall conductance
as the gap is opened,
$\delta \sigma_{xy}=\sigma_{xy}(\Delta)-\sigma_{xy}(\Delta=0)$.
For small $\Delta$ the important changes
in the Fermi surface occur near
the points where the $\Delta=0$ Fermi surface
crosses its translation by ${\bf Q}$,
{\it i.e.} where $\varepsilon_p=\varepsilon_{p+Q}=\mu$.
One of these crossing points is shown in Fig \ref{fig:localview}.
The change of chemical potential is second order
effect of $\Delta$, and thus neglected in our analysis.
To simplify notation,
in this section, we measure $\mu$ from its value at criticality.
Thus the SDW Fermi surface with backscattering potential $\Delta$,
$E^{\pm}_p=0$,
can be written as $\varepsilon_p\varepsilon_{p+Q}=\Delta^2$.
Any change in the scattering rate is similarly of order $\Delta^2$
and is neglected. Near such
a crossing point we may write $\delta \sigma_{xy}$
in terms of the contributions
from the two SDW bands $E^{\pm}_p$
and the original Fermi surface, $p$,
and its image under translation by ${\bf Q}$, $sp$, as
\begin{multline}
\delta \sigma_{xy}\sim \int_{-\infty}^{\infty} ds ds'
\biggl( \sum_{b=\pm} K_b(s,s')\sin[\phi_b(s)-\phi_b(s')] \\
   -\sum_{a=p,sp} K_a(s,s')\sin[\phi_a(s)-\phi_a(s')] \biggr)
\label{deltasig}
\end{multline}
We shall see that this integral is dominated
by $s,s'$ close to the crossing
point, with ``close'' determined by the larger
of $a$ (evaluated at the crossing point)
and $p_{\Delta}$, so that we have neglected the fact
that the Fermi surface closes
on itself and have extended the integration to infinity.

We analyse Eq [\ref{deltasig}]
by expanding the energy dispersions about the crossing point.
In the first approximation, one may consider taking $\tau$
to be constant and linearizing the dispersions
about the crossing point,
writing $\varepsilon_p={\bf v}^p\cdot {\bf \delta p}$
and  $\varepsilon_{p+Q}={\bf v}^{sp}\cdot {\bf \delta p}$,
with ${\bf v}^{p,sp}$ constant. In this approximation,
the $\Delta=0$ Fermi surface and its image are straight lines,
so $\phi^{p,sp}$ are independent
of $s$, so the $\Delta=0$ contribution vanishes,
and the two SDW bands
are mirror images of each other with respect to
the dot-dashed line in Fig \ref{fig:localview},
so $K_{+}(s)=K_-(s)$ and
$\phi_+(s)-\phi_+(s')=-(\phi_-(s)-\phi_-(s'))$,
so that $\delta \sigma_{xy}$ vanishes.
Obtaining a non-zero $\delta\sigma_{xy}$ requires
consideration of effects which break this mirror symmetry.
These include variation of $\tau$ or $|v^p|$ with momentum
as well as curvature of the $\Delta=0$ Fermi surface.
Formally, all of these effects are of order $p_0^{-1}$,
and we work to leading order in this quantity. Then we may write
\begin{equation}
a^{\pm}(s)=a(s)\left[1\pm \delta a(s)\right]
\end{equation}
where $a(s)$ is the corresponding function
in presence of mirror symmetry,
and $\delta a(s)$ represents the broken mirror symmetry,
and thus is of order $\mathcal{O}(1/p_0)$.
In terms of these functions:
\begin{equation}
\begin{split}
K_{\pm}(s,s^{\prime}) &\approx K(s,s^{\prime})
\left(1\pm\int_{s^{\prime}}^s dx
\frac{\delta a(x)}{a(x)}\right)\\
 &=K(s,s^{\prime})(1\pm \delta K)
\end{split}
\end{equation}
We also have:
\begin{equation}
\phi_+(s)-\phi_+(s^{\prime})+\phi_-(s)-\phi_-(s^{\prime})
=f(s,s^{\prime})
\end{equation}
where $f$ is also of order $1/p_0$.
Thus, Eq [\ref{deltasig}] can be expanded as:
\begin{equation}\label{criticaldeltasig}
\begin{split}
\delta \sigma_{xy}\sim\int ds ds^{\prime}
\Bigl\{&2K(s,s^{\prime})f(s,s^{\prime})
\cos[\phi_+(s)-\phi_+(s^{\prime})]\\
 &+ 2K(s,s^{\prime})\delta K(s,s^{\prime})
\sin[\phi_{+}(s)-\phi_+(s^{\prime})]\\
 &- K_{p}(s,s^{\prime})g(s,s^{\prime})
-K_{sp}(s,s^{\prime})g(s,s^{\prime})\Bigr\}
\end{split}
\end{equation}
with $g(s,s^{\prime})$ of order $1/p_0$.
In obtaining this equation, we have noticed that
$f(s,s^{\prime})$, $\delta K$ and $g(s,s^{\prime})$
are already of order $1/p_0$,
so that all the remaining quantities
can be taken from their unperturbed counterparts.
We also notice that $a^{p}(s)=a^{sp}(s)$,
and in this approximation they are constant,
denoted by $a^p$ in the following discussion.

The detailed analysis of this equation is somewhat invovled,
except for the strong magnetic field limit to be discussed below,
partly because it is not easy to obtain an analytic relationship
between the arclength coordinates $s$
and momentum coordinate-{\bf p} for SDW Fermi line.
We shall argue that this $\delta \sigma_{xy}$ has a scaling form:
\begin{equation}
\delta\sigma_{xy}=C(B\Delta)F(a^p/p_{\Delta})
\label{scalingfunction}
\end{equation}
such that $F(0)=1$ and $F(x\rightarrow\infty)\propto x$,
with the prefactor $C$ of order $1/p_0$,
which will be computed in the special case of
constant relaxation time in the following,
and the exact form of $F$ depending on
the manner in which the symmetry
between $+$ and $-$ branches is broken.

It has been noted by Bazaliy {\it et al} \cite{Bazaliy04}
that in the vicinity of
the crossing points, ${\bf p}$
can be parametrized by $\varepsilon_{p}$ and
$\varepsilon_{p+Q}$:
first expanding the dispersions
\begin{equation}\label{expansion}
\begin{split}
\varepsilon_p&={\bf v}^{p\star}\cdot\delta{\bf p}+\frac{1}{2}m_{ij}
\delta p_{i}\delta p_{j}+O(\delta p^{3})\\
\varepsilon_{p+Q}&={\bf v}^{sp\star}\cdot\delta{\bf p}+
\frac{1}{2}n_{ij} \delta p_{i}\delta p_{j}+O(\delta p^{3})
\end{split}
\end{equation}
where, $\delta{\bf p}={\bf p}-{\bf p}^{\star}$,
${\bf v}^{p,sp\star}={\bf v}^{p,sp}({\bf p}^{\star})
\equiv (\nabla_p\varepsilon_{p,p+Q})^{\star}$,
$m_{ij}=(\partial^{2} \varepsilon_p/
\partial p_{i}\partial p_{j})^{\star}$,
$n_{ij}=(\partial^{2} \varepsilon_{p+Q}/
\partial p_{i}\partial p_{j})^{\star}$,
and the superscript $\star$ stands for
the corresponding value evaluated at the crossing point,
and then inverting them
\begin{equation}
\delta{\bf p}={\bf u}_{1}\varepsilon_p+{\bf u}_{2}\varepsilon_{p+Q}
=({\bf u}_{1}+\frac{\Delta^{2}}{\varepsilon_p^{2}}{\bf u}_{2})
\varepsilon_p
\label{inverse}
\end{equation}
with
${\bf u}_{1}=\frac{{\bf v}^{sp\star}\times
[{\bf v}^{p\star}\times{\bf v}^{sp\star}]}
{({\bf v}^{p\star}\times{\bf v}^{sp\star})^2}$
and
${\bf u}_{2}=\frac{{\bf v}^{p\star}\times
[{\bf v}^{sp\star}\times{\bf v}^{p\star}]}
{({\bf v}^{p\star}\times{\bf v}^{sp\star})^2}$.
From Eq [\ref{inverse}]:
\begin{equation}
d{\bf p}={\bf u}_{1}d\varepsilon_p+{\bf u}_{2}d\varepsilon_{p+Q} =
({\bf u}_{1}-\frac{\Delta^{2}}{\epsilon_p^{2}}{\bf u}_{2})d\epsilon_p
\label{dp}
\end{equation}
where the second equality of
Eqs [\ref{inverse},\ref{dp}] only holds on the
SDW Fermi surface. Thus,
\begin{multline}
\left|\frac{ds}{d\varepsilon_p}\right|=\frac{1}
{|{\bf v}^p\times{\bf v}^{sp}|}
\Bigl\{(v^{sp})^2+(v^p)^2\left(\frac{\Delta^2}
{\varepsilon_p^2}\right)^2\\+
2\left(\frac{\Delta^2}{\varepsilon_p^2}\right)
{\bf v}^{p}\cdot{\bf v}^{sp}\Bigr\}^{1/2}
\label{convertds}
\end{multline}
And the SDW Fermi velocities are:
\begin{equation}
{\bf v}^{\pm}({\bf p})=\frac{\Delta^2}{\varepsilon_p^2+\Delta^2}
{\bf v}^p({\bf p})+\frac{\varepsilon_{p}^2}
{\varepsilon_{p}^2+\Delta^{2}}{\bf v}^{sp}({\bf p})
\label{SDWvelocity}
\end{equation}
Now return to Eq [\ref{criticaldeltasig}].
$a(s)=\pi B\tau_0v(s)/\Phi_0$,
where $\tau_0$ is the relaxation time at crossing point,
$v(s)$ is the SDW Fermi velocity when mirror symmetry is present,
in which case ${\bf v}^p$ and ${\bf v}^{sp}$ in
Eqs [\ref{convertds},\ref{SDWvelocity}] are constant vectors.
Substituting this and Eq [\ref{convertds}]
into Eq [\ref{criticaldeltasig}],
scaling energy variables in unit of $\Delta$,
and defining
$\mathcal{B}=B(v_F^2\pi\tau_0\sin\Delta\phi/\Phi_0)$,
where $v_F=v^{p,sp}$ and $\Delta\phi$ is the angle
made by the normal state Fermi velocities,
${\bf v}^p$ and ${\bf v}^{sp}$,
shown in Fig \ref{fig:localview},
we can obtain the above scaling form Eq [\ref{scalingfunction}].
In the following, we shall discuss
Eq [\ref{scalingfunction}] in two limits:
weak magnetic field, $a^p\ll p_{\Delta}$,
and strong magnetic field, $a^p\gg p_{\Delta}$.

\begin{figure}
 \centering
 \includegraphics[width=3.0in]{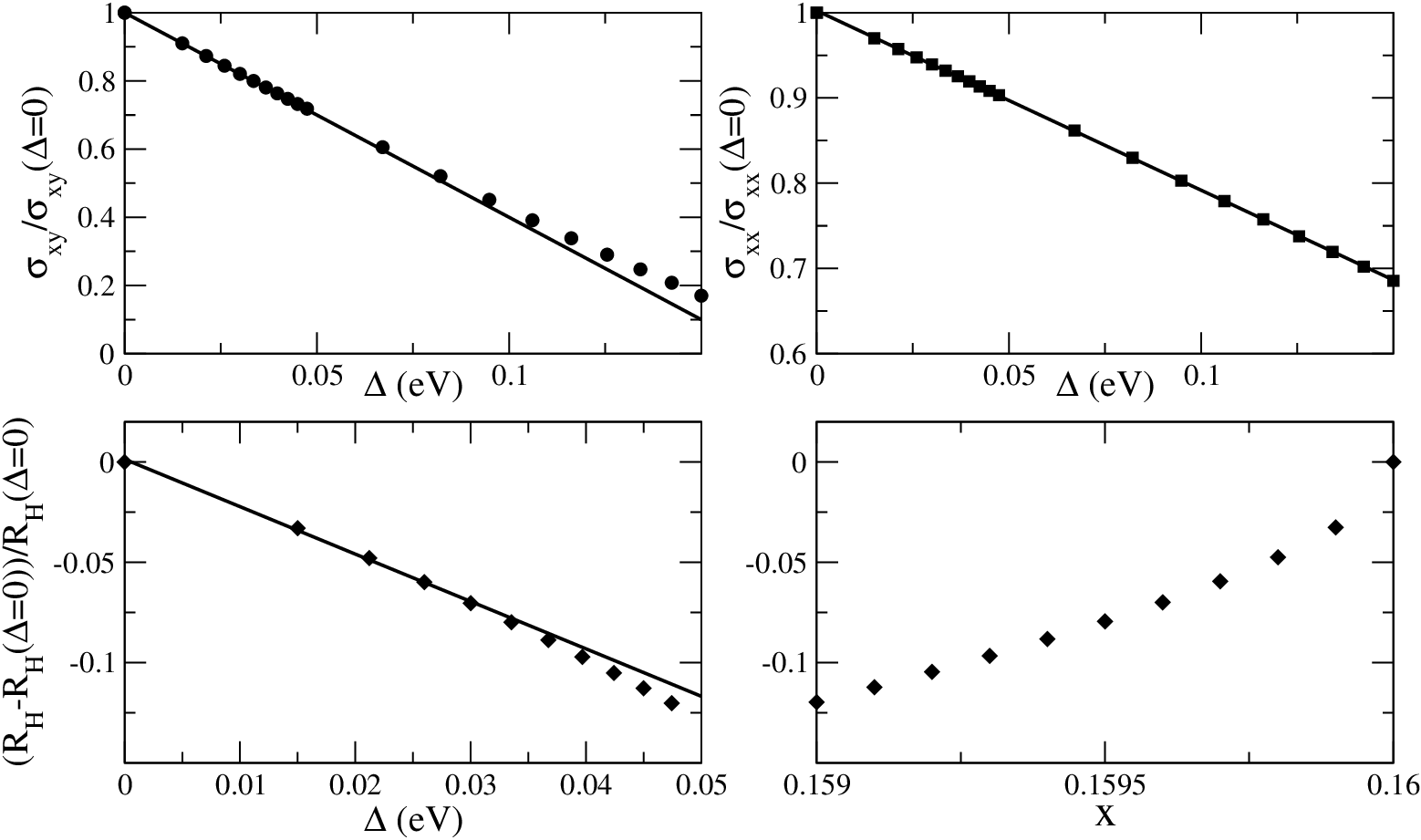}
 \caption{\textit{Upper left: numerical results for
 $\sigma_{xy}(\Delta)/\sigma_{xy}(\Delta=0)$ as a function of
 $\Delta$ near critical doping (solid circle)
and function $1.0-6.1 \Delta$ (solid line).
Upper right: numerical results for
$\sigma_{xx}(\Delta)/\sigma_{xx}(\Delta=0)$
as function of $\Delta$ (solid square)
and function $1.0-2.1\Delta$ (solid line).
Lower left: numerical results for $R_H$
as a function of $\Delta$ (solid diamond)
and function $-2.4\Delta$ (solid line).
Lower right: numerical results for $R_H$
as a function of $x$ near critical doping.
The parameters are: $t_1=0.38$eV, $t_2=0.32t_1$, $t_3=0.075$eV,
and $\Delta(x)[eV]=0.6\sqrt{1-x/0.16}$.}}
\label{fig:linearbehavior}
\end{figure}

In weak magnetic field, we can start from Eq [\ref{ongformula}].
Both ${\bf v}^p$ and $\tau$ can vary in momentum space.
To lowest order, these effects are additive.
If we keep the {\bf p}-dependence of relaxation time $\tau$,
assuming its dependence has form:
\begin{equation}
\tau_p=\tau_0\left[1+\mathcal{A}
(\varepsilon_p+\varepsilon_{p+Q})\right]
\label{relaxationfunction}
\end{equation}
and keep velocity of normal state,
${\bf v}^{p}$ and ${\bf v}^{sp}$, to be constant,
Eq [\ref{ongformula}] can be evaluated for each Fermi pocket,
giving
\begin{equation}
\delta \sigma_{xy}=\sigma_Q4\tau_0^2\mathcal{A}
\frac{B\Delta}{\Phi_{0}}
{\hat {\bf z}}\cdot[{\bf v}^{p}\times{\bf v}^{sp}]
\label{weakfieldtau}
\end{equation}
In the case in which the normal state velocities
${\bf v}^{p}$ and ${\bf v}^{sp}$ are {\bf p}-dependent
and relaxation time $\tau$ is constant,
differentiating Eq [\ref{expansion}],
substituting Eq [\ref{inverse}] into it,
and defining vectors:
\begin{align*}
\vec\eta^p_{1} &= (m_{11}u_{1x}+m_{12}u_{1y})\hat{{\bf x}}
+(m_{21}u_{1x}+m_{22}u_{1y})\hat{{\bf y}}\\
\vec\eta^p_{2} &= (m_{11}u_{2x}+m_{12}u_{2y})\hat{{\bf x}}
+(m_{21}u_{2x}+m_{22}u_{2y})\hat{{\bf y}}\\
\vec\eta^{sp}_{1} &= (n_{11}u_{1x}+n_{12}u_{1y})\hat{{\bf x}}
+(n_{21}u_{1x}+n_{22}u_{1y})\hat{{\bf y}}\\
\vec\eta^{sp}_{2} &= (n_{11}u_{2x}+n_{12}u_{2y})\hat{{\bf x}}
+(n_{21}u_{2x}+n_{22}u_{2y})\hat{{\bf y}}
\end{align*}
we then have
${\bf v}^b={\bf v}^{b\star}+\delta {\bf v}^{b}
={\bf v}^{b\star}+(\vec\eta^b_{1}+
\vec\eta^b_{2}(\Delta/\varepsilon_{p})^{2})\varepsilon_p$
and
$d{\bf v}^{b}=(\vec \eta^b_{1}-\vec \eta^b_{2}
(\Delta/\varepsilon_{p})^{2})d\varepsilon_p$, ($b=p,sp$).
Eq [\ref{ongformula}] then gives:
\begin{equation}
\delta \sigma_{xy}=\sigma_Q\tau^2\frac{B\Delta}{\Phi_0}
{\bf \hat{z}}\cdot[(\vec \eta^p_{1}+
\vec \eta^{sp}_{2}+3\vec \eta^p_{2}
+3\vec \eta^{sp}_{1})
\times({\bf v}^{sp\star}-{\bf v}^{p\star})]
\label{weakfieldcurvature}
\end{equation}
These weak magnetic field results,
Eqs [\ref{weakfieldtau},\ref{weakfieldcurvature}],
confirm our scaling form for $\delta\sigma_{xy}$,
Eq [\ref{scalingfunction}],
in the limit $a^p/p_{\Delta}\ll 1$.
Also the prefactor $C$ can be calculated
from these two equations.
It would be of interest to study
Eq [\ref{weakfieldcurvature}] in more detail,
since all the parameters in
$\delta\sigma_{xy}/\sigma_{xy}(x=x_c+)$ are set by
normal state energy function Eq [\ref{dispersion}].
Taking $\sigma_{xy}(x=x_c+)$ from numerical calculation,
$\delta\sigma_{xy}/\sigma_{xy}(x=x_c+)=-6.1\Delta$,
shown as solid line in the upper left panel of
Fig \ref{fig:linearbehavior},
which is compared with the results of numerical evaluation
of Eq [\ref{ongformula}] (solid circles in this panel).
In weak magnetic field,
the longitudinal conductivity $\sigma_{xx}$
has been treated in great detail for the case of
the 3d SDW transition in $Cr$. \cite{Bazaliy04}
Following essentially the same procedure,
we obtain a similar expression for $\delta\sigma_{xx}$:
\begin{equation}
\delta \sigma_{xx}
=-\sigma_Q\frac{\tau}{\pi}\frac{({\bf v}^{p\star}-
{\bf v}^{sp\star})^{2}}{|{\bf v}^{p\star}\times{\bf v}^{sp\star}|}
\Delta
\label{weakfieldsigmaxx}
\end{equation}
Taking $\sigma_{xx}(x=x_c+)$ from numerical calculation,
$\delta\sigma_{xx}/\sigma_{xx}(x=x_c+)=-2.1\Delta$,
shown as solid line in upper right panel of
Fig \ref{fig:linearbehavior},
compared with the results taken from numerical evaluation
of Eq [\ref{sigxxlow}] (solid squares).
The numerical results of
$R_H\equiv\sigma_{xy}/(B\sigma_{xx}^2)$ are shown as solid diamonds
in the lower panels of Fig \ref{fig:linearbehavior}.
The reconnection of Fermi surface suppresses
both longitudinal and transverse conductivity.
In the vicinity of criticality,
$R_{H}/R_{H}(x=x_c+)\approx
1+\delta\sigma_{xy}/\sigma_{xy}(x=x_c+)
-2\delta\sigma_{xx}/\sigma_{xx}(x=x_c+)\approx 1.0-1.9\Delta$.
Since $\Delta\sim\sqrt{x_{c}-x}$,
we expect a square root behavior in
$R_{H}(x)$. There actually is, as shown in the lower
right plot of Fig \ref{fig:linearbehavior},
however, this regime is so
small that it may be difficult to observe in practice.

\begin{figure}
\centering
\includegraphics[width=2.5in]{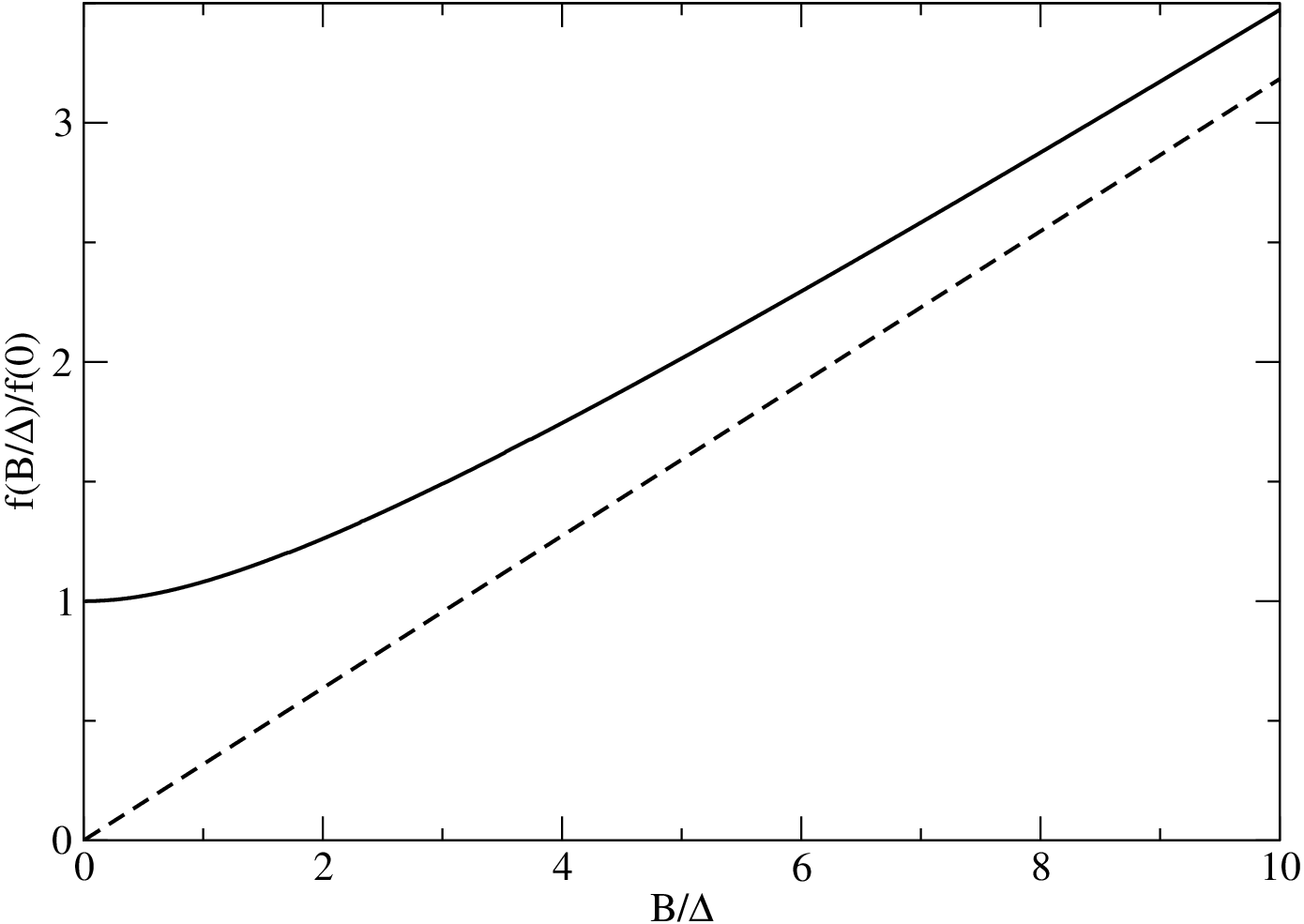}
\caption{\textit{The scaling function $f(B/\Delta)$ for
$\delta\sigma_{xy}$ under the linearized dispersion,
constant velocity approximation.
The unit of $B/\Delta$ is $\mathcal{B}/\Delta$.}}
\label{fig:vary_relaxationtime}
\end{figure}

In strong magnetic field regime $a^p/p_{\Delta}\gg 1$,
the integration Eq [\ref{criticaldeltasig}] is
dominated by ($s>0,s^{\prime}<0$), with
$f(s,s^{\prime}),\,\,\,g(s,s^{\prime})\sim (s-s^{\prime})/p_0$,
$\delta a(s)\sim {\rm sgn}s \cdot s/p_0$,
and $\phi_+(s)-\phi_+(s^{\prime})\approx\Delta\phi$, then
\begin{equation}
\delta \sigma_{xy}=\sigma_Q\frac{\left(l/l_B\right)^3}{\pi^2l_Bp_0}
[\cos\Delta \phi-1+\alpha \sin \Delta\phi]
\label{strongfieldresult}
\end{equation}
with $\alpha\sim \frac{p_0}{l}\frac{\partial l}{\partial s}\sim 1$
and $l$ the mean free path in normal state.
This result, $\delta \sigma_{xy}\sim B^2$
and independent of $\Delta$,
confirms our scaling function for $a^p/p_{\Delta}\gg 1$.
To justify this result, let's consider a simple case
in which the normal state velocities ${\bf v}^{p,sp}$ are constant,
while relaxation time is assumed to be of form:
Eq [\ref{relaxationfunction}].
Under this condition, the only contribution to
 $\delta \sigma_{xy}$ in Eq [\ref{criticaldeltasig}]
comes from the second line containing $\delta K$.
Changing to $\varepsilon_p$ variable, it gives
\begin{equation}
\delta \sigma_{xy}=\sigma_Q\frac{\tau_0\mathcal{A}}
{4\pi^2}\mathcal{B}\Delta f(\mathcal{B}/\Delta)
\label{scalingrelaxation}
\end{equation}
$f(x)$ is a function of only one variable,
numerical evaluation of which gives the solid line
in Fig \ref{fig:vary_relaxationtime},
along which shown as dashed line is the linear function $f(x)=2x$,
representing the calculated asymptotic behavior
as $x\rightarrow \infty$.
The prefactor 2 is calculated by relating $\alpha$ in
Eq [\ref{strongfieldresult}] and
$\mathcal{A}$ in Eq [\ref{relaxationfunction}] in the limit
$\Delta\rightarrow 0$: $\alpha/p_0=\mathcal{A}(v_F\sin \Delta\phi)/2$.
This solid line clearly shows the expected behavior of $f(x)$: $f(x)=1$
for $x\rightarrow 0$, and $f(x\rightarrow \infty)\propto x$,
with a crossover in between.
The crossover regime can be estimated by setting $\mathcal{B}=\Delta$.
The relaxation time can be inferred from the {\it ab}
plane resistivity measurement: \cite{Dagan04} at B=10 T,
$\rho\sim 20 \mu\Omega\cdot{\rm cm}$,
which gives $\tau\sim 10^{-14}$ sec.
So, for $\Delta=0.1$eV, corresponding to $x_c-x\sim 0.005$
if we take $\Delta(x)=0.7\sqrt{1-x/x_c}$eV,
the crossover is $B\sim \Phi_0\hbar\Delta/(v_F^2\tau)\sim 100$T,
where $v_F\approx 4$eV$\cdot$\AA.
For these parameters, the breakdown field is slightly larger.

\section{Sensitivity to Parameters}
\label{sec:Sensitivity}

In this section we examine the sensitivity of our results
to different choices of parameters.

\begin{figure}
\centering
\includegraphics[width=2.5in]{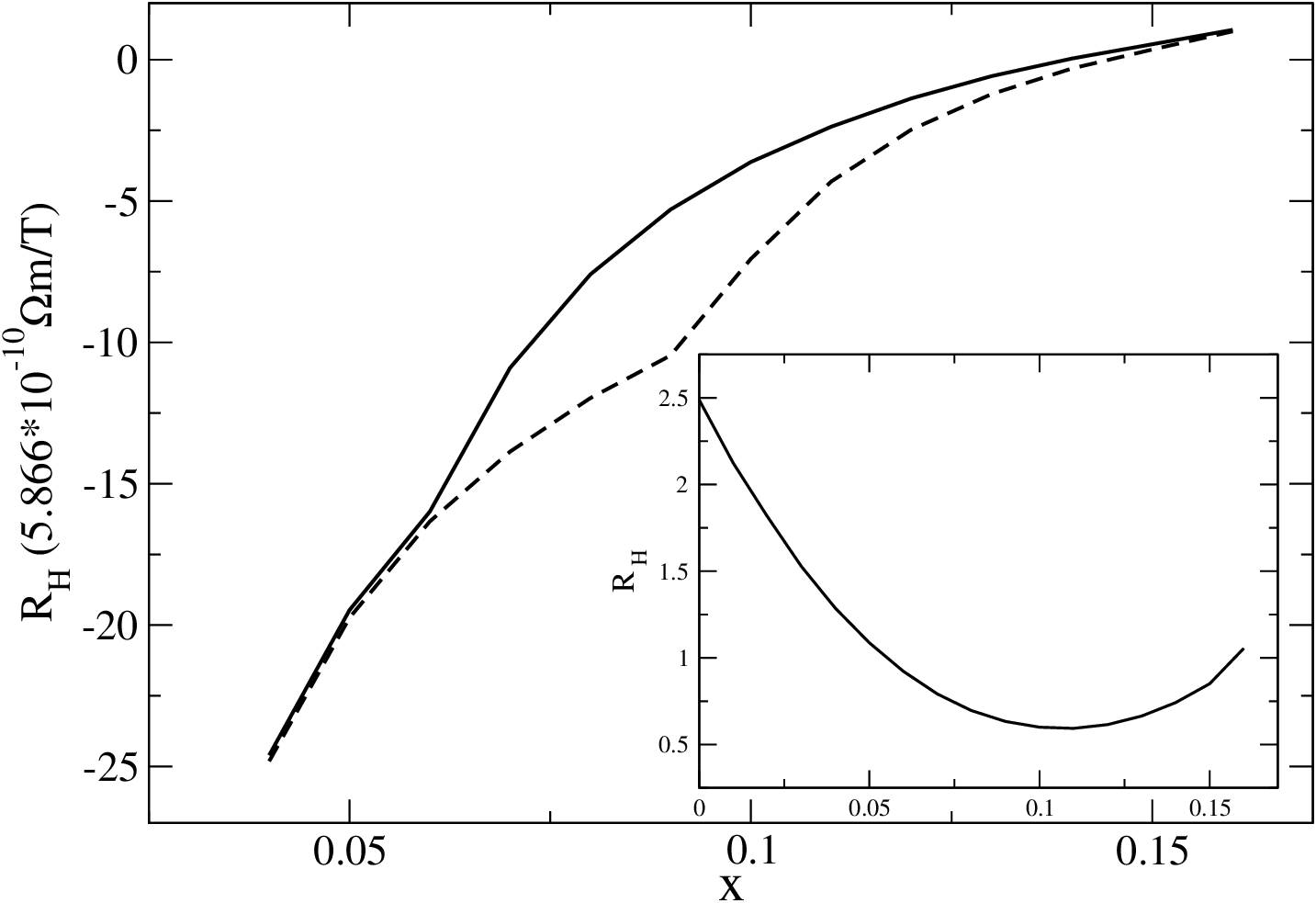}
\caption{\textit{Numerical results of Hall coefficient
in SDW ordered state for different choices of parameters.
Dashed line: $t_1=0.38$eV, $t_2=0.32t_1$, $t_3=0.075$eV
and $\Delta(x=0)=0.6$eV. Solid line:
$t_3=0.5t_2$, $t_2=0.32t_1$, $\Delta(x=0)=0.5$eV.
Inset: $R_H$ for $\Delta(x=0)=0.3$eV, with the same
band parameters as the solid line in the main panel.}}
\label{fig:compareorderedhall}
\end{figure}

Let us first consider the $x<x_{c}$ region.
Fig \ref{fig:compareorderedhall} shows numerical results
of $R_H$ for parameter choices:
$t_1=0.38$eV, $t_2=0.32t_1$, $t_3=0.075$eV
and $\Delta(x=0)=0.6$eV (dashed line),
and $t_2=0.32t_1$, $t_3=0.5t_2$
and $\Delta(x=0)=0.5$eV (solid line),
which can be compared with the solid and dashed lines
in Fig \ref{fig:hall}.
All of these curves share the same feature:
the existence of $x_1$.
In Fig \ref{fig:hall}, $x_1\approx 0.12$,
while in Fig \ref{fig:compareorderedhall},
$x_1\approx 0.09$ (dashed line)
and $x_1\approx 0.06$ (solid line).
For $x>x_1$, both electron pocket (centered at $(0,\pi)$) and
hole pockets (centered at $(\pi/2,\pm\pi/2)$) are present, and as
doping decreases, both of them contribute to the decrease of $R_{H}$,
while for $x<x_1$, the hole pockets vanish (the $E_p^-$
band is filled), and thus only electron pocket contributes to $R_{H}$,
which leads to the slowing down of decrease
in $R_{H}$ as decreasing doping.
For $x<x_1$, the behavior of $R_{H}$ is quite universal, approximately
$\sim -1/x$ (the Hall coefficient for free electrons,
with $x$ electron density), as shown in Fig \ref{fig:hall}.
Different choices of parameters
change the actual value of $x_1$. The  numerical amplitude
of $R_{H}(x_1)$ is approximately $1/x_1$, by continuity.
There is still another possibility that
if the opening of SDW gap is not fast
enough, the hole pockets survive for $x$ down to 0. In this
case, there is an upturn when $x$ is smaller than some value,
with a much smaller relative change in the magnitude of $R_H$,
as shown in the inset of Fig \ref{fig:compareorderedhall}.
We do not believe this possibility is relevant
to the electron doped cuprates.

\begin{figure}
\centering
\includegraphics[width=2.5in]{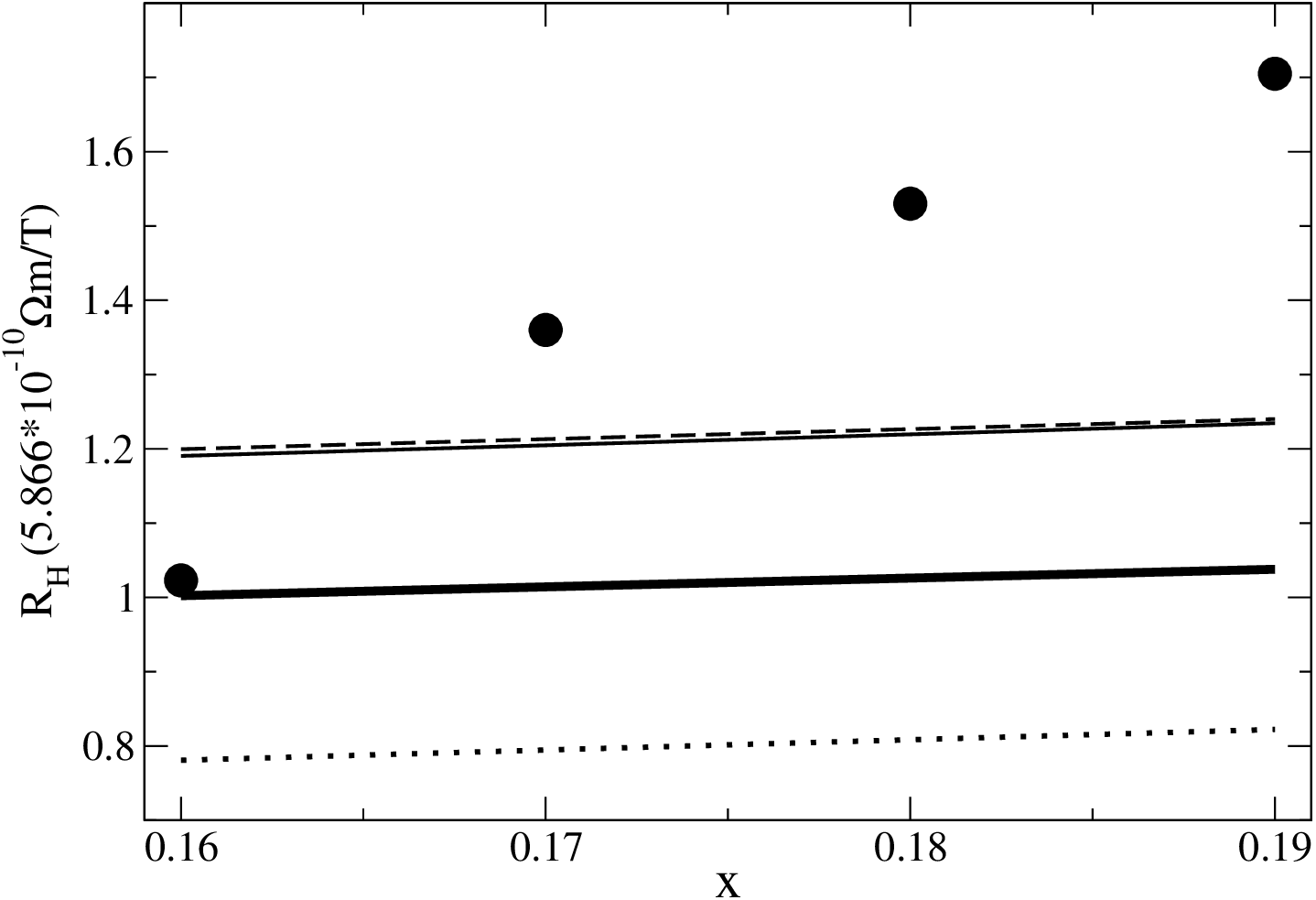}
\caption{\textit{Numerical results of Hall coefficient
in normal state for different
choices of band parameters. Heavy solid line: $t_2=0.32t_1$ and
$t_3=0.075$eV; dashed line:  $t_2=0.2t_1$ and $t_3=0.5t_2$;
dotted line: $t_2=0.45t_1$ and $t_3=0.5t_2$; light solid line:
curve $1/(1-x)$;
solid circles: experimental data. \cite{Dagan04}}}
\label{fig:comparenormalhall}
\end{figure}

\begin{figure}[htbp]
\centering
\includegraphics[width=2.5in]{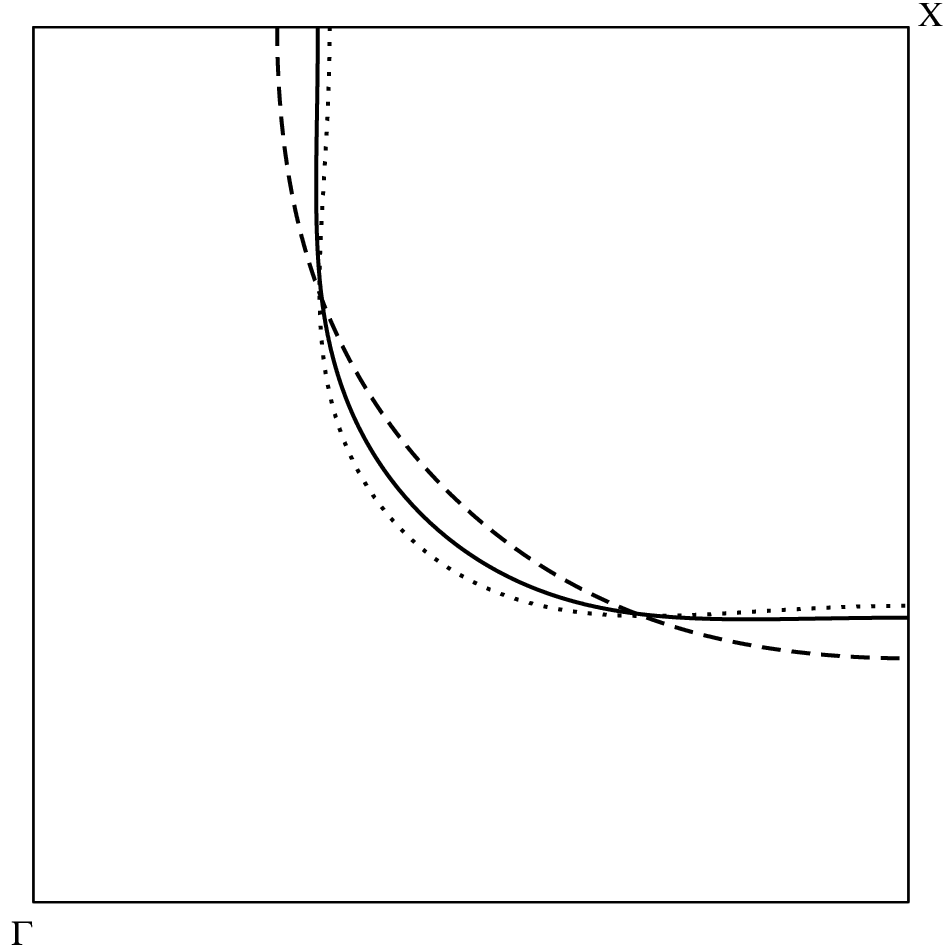}
\caption{\textit{Normal state Fermi surfaces
in the first quadrant of Brillouin zone
for different choices of band parameters at $x=0.16$.
Solid line: $t_2=0.32t_1$ and $t_3=0.075$eV;
dashed line:  $t_2=0.2t_1$ and $t_3=0.5t_2$;
dotted line: $t_2=0.45t_1$ and $t_3=0.5t_2$.}}
\label{fig:comparenormalfs}
\end{figure}

In the regime $x>x_c$, we can change band parameters
$t_2$ and $t_3$ in Eq [\ref{dispersion}],
and calculate $R_H$ for these choices of
parameters. Fig \ref{fig:comparenormalhall}
shows the results for different
choices of band parameters: $t_2=0.32t_1$
and $t_3=0.075$eV (heavy solid line);
$t_2=0.2t_1$ and $t_3=0.5t_2$ (dashed line);
$t_2=0.45t_1$ and $t_3=0.5t_2$ (dotted line).
The Fermi surfaces corresponding to these choices
of parameters at $x=0.16$
are shown in Fig \ref{fig:comparenormalfs}.
It can be seen that
(i) different choices of band parameters
do not lead to drastic difference in $R_H$;
(ii) for each choice of band parameters,
$R_{H}$ is quite insensitive
to change in doping, in contrast to the experimental data,
shown as solid circles;
(iii) our numerical $R_H$ are close to $1/(1-x)$
(the Hall coefficient for circular Fermi surface,
with $1-x$ hole density),
shown as light solid line in
Fig \ref{fig:comparenormalhall}. Both
our numerical results and $1/(1-x)$ are not close to
the experimental data of Ref. \onlinecite{Dagan04}.
The difference is largest at $x=0.19$,
the point farthest from criticality,
where one would expect the difference to be smallest.

\section{$x>x_c$}
\label{sec:Normal State Region}

Fig \ref{fig:comparenormalhall} shows
that the measured $R_H$ at $x>x_c$ shows a pronounced
doping dependence which is incompatible
with the mean field calculation at fixed band parameters.
Fig \ref{fig:comparenormalhall} also shows that a variation
of the correct order of magnitude can be obtained
if a doping dependence of the band parameters is assumed.
In this section we investigate whether this doping dependence
can be viewed as a precursor of the spin density wave instability
by calculating the effect of spin fluctuations
on the Fermi surface. This is given
by the real part of the self energy.
The leading approximation to this is:
\begin{equation}
\Sigma({\bf k},i\omega)=-g^{2}T\int (d{\bf q})\sum_{i\Omega_{n}}
\mathcal{G}({\bf k}+{\bf q},i\omega+i\Omega_{n})
\mathcal{D}({\bf q},i\Omega_{n})
\label{selfenergy}
\end{equation}
where $\mathcal{G}$ and $\mathcal{D}$ are
electron and spin fluctuation
Matsubara Green's functions, respectively.
In our calculation, we shall take:
$\mathcal{G}({\bf p},i\omega)=(i\omega-\varepsilon_p)^{-1}$ and
$\mathcal{D}({\bf q},i\Omega_n)=-(\Gamma_q+|\Omega_n|)^{-1}$,
with $\Gamma_q=\Gamma(r+\xi_0^2({\bf q}-{\bf Q})^2)$, where
$\Gamma$ is the energy scale characteristic of spin fluctuations,
 $\xi_0$ is coherent length,
$r$ characterizes the distance to criticality,
and ${\bf Q}=(\pi,\pi)$.

This self energy for $i\omega=0$ can be integrated immediately,
at $T\rightarrow 0$,
if both $\varepsilon_{k+Q}/E_0$ ($E_0$ is an energy cutoff)
and $r$ are small.
The result may be represented as
\begin{equation}
\Sigma({\bf k},i\omega=0)=-0.5\lambda\varepsilon_{k+Q}
\ln\frac{1}{(r^{2}+(\varepsilon_{k+Q}/E_{0})^{4})}
\label{fsrenormalization}
\end{equation}
where, $\lambda$ is a dimensionless constant.

\begin{figure}[htbp]
\centering
\includegraphics[width=2.5in]{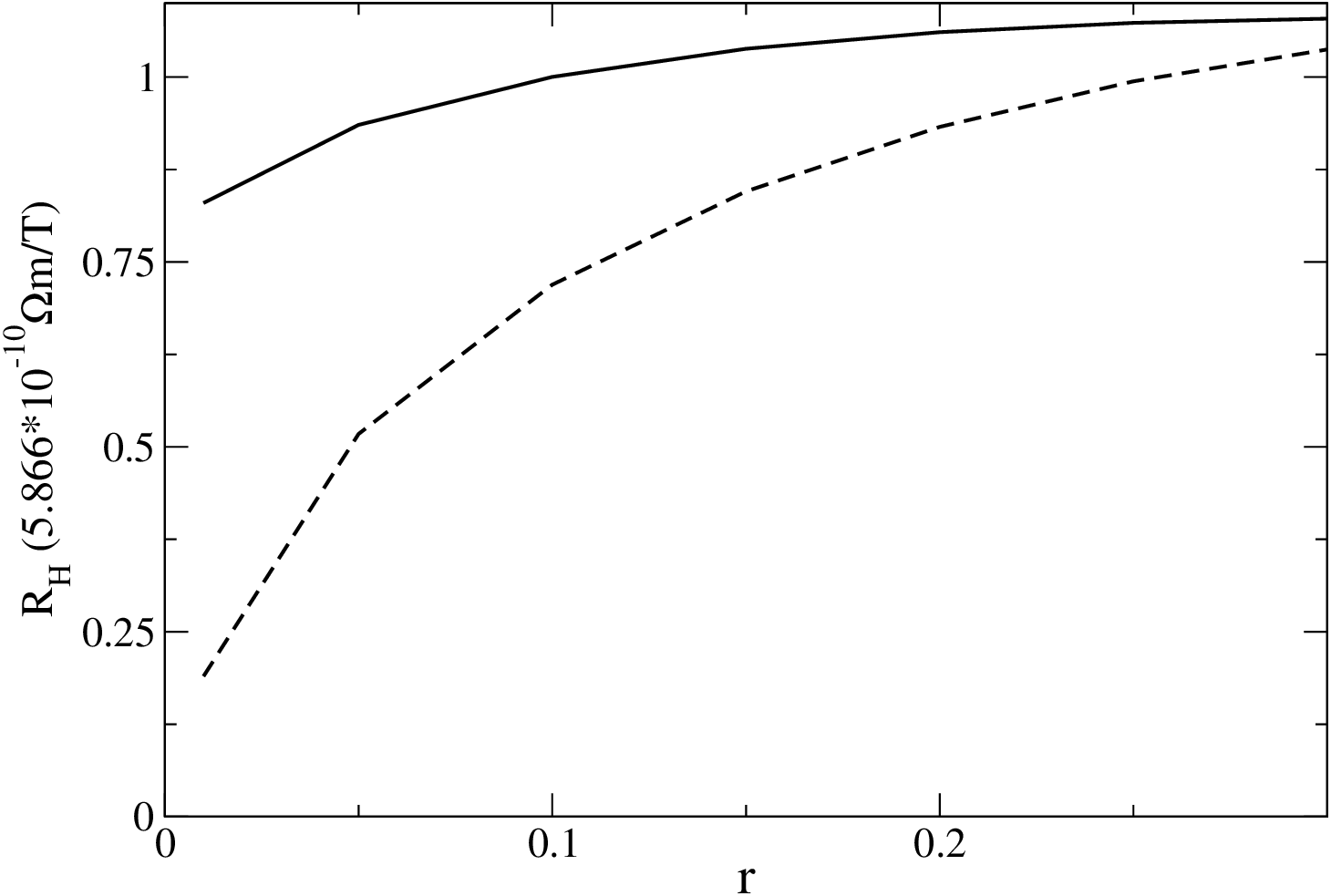}
\caption{\textit{Numerical result for Hall coefficient
as a function of $r$
when electronic energy dispersion is
$\varepsilon_p+\Sigma({\bf p},i\omega=0)$,
for coupling constant $\lambda=0.2$ (solid line)
and $\lambda=0.4$ (dashed line).
Other parameters are: $t_1=0.38$eV,
$t_2=0.5t_1$, $t_3=0.075$eV,
the energy cutoff $E_0=0.88$eV.}}
\label{fig:fsrenormalhall}
\end{figure}

\begin{figure}[htbp]
\centering
\includegraphics[height=2.5in]{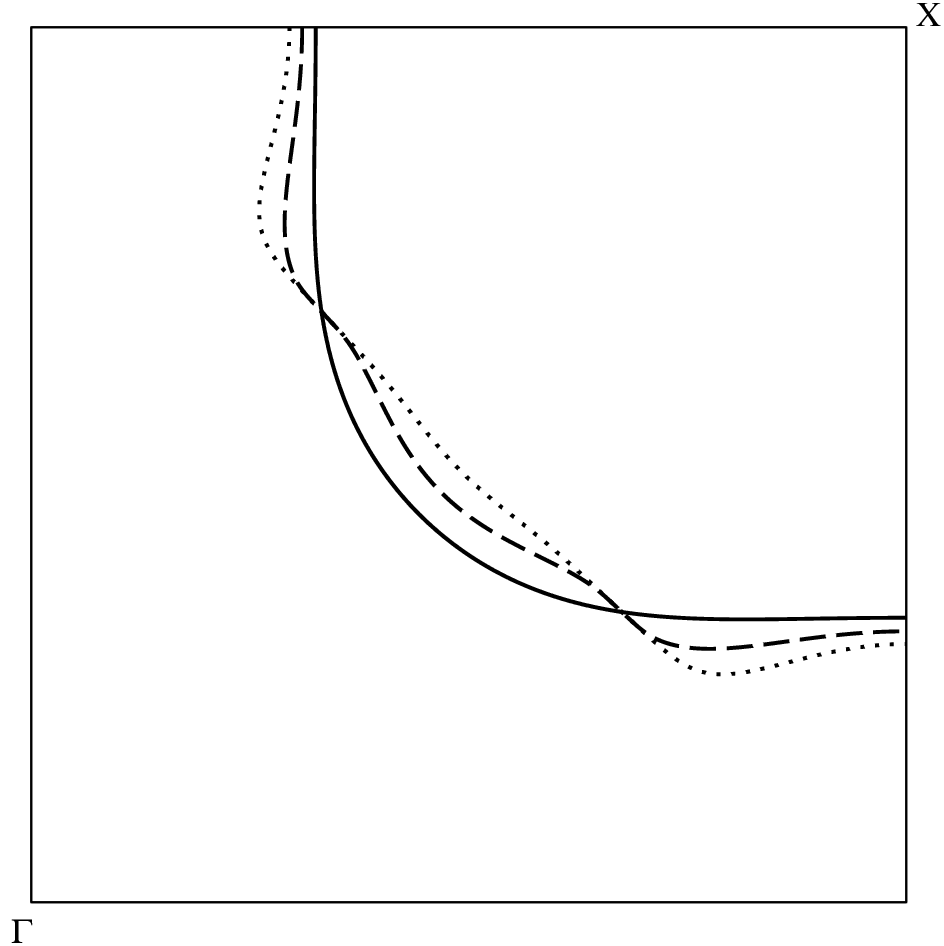}
\caption{\textit{Fermi surfaces in normal state,
with renormalized energy
dispersion relation, Eq [\ref{fsrenormalization}]:
$\lambda=0$ (solid line);
$\lambda=0.2$ and $r=0.01$ (dashed line);
$\lambda=0.4$ and $r=0.1$ (dotted line). The numerical
$R_H$ corresponding to the dashed line is about
20\% less than that corresponding to the solid line,
and $R_H$ corresponding to dotted line
is about 30\% less than that of the solid line.}}
\label{fig:renormalizedfs}
\end{figure}

We can take $\varepsilon_p+\Sigma({\bf p},i\omega=0)$ as our new
electronic energy dispersion relation,
and integrate Eqs [\ref{sigxxlow},\ref{ongformula}] numerically.
The Hall coefficient $R_H$ as function of $r$,
for $\lambda=0.2$ (solid line) and $\lambda=0.4$ (dashed line),
is shown in Fig \ref{fig:fsrenormalhall},
where the parameters are taken as: $t_1=0.38$eV,
$t_2=0.5t_1$, $t_3=0.075$eV, and $E_0=0.88$eV.
Because the relationship between $r$ and
doping has not yet established, in our numerical calculation,
we keep the chemical potential fixed
at the unrenormalized value for $x=0.16$.
Fig \ref{fig:renormalizedfs} shows Fermi surfaces
for the case $\lambda=0.2$, $r=0.01$
and the case $\lambda=0.4$, $r=0.1$,
as well as the original unrenormalized one.
Both of these Fermi surfaces give $R_H$ about
25\% less than the unrenormalized value.

\section{Conclusion: discussion of experiment}
\label{sec:Conclusion and Discussion}

We have used mean field theory to calculate the changes
in the Hall resistivity of a two dimensional system
which undergoes a spin density wave instability.
The model studied was chosen to be appropriate
to the electron-doped cuprates.
The important physics is that as $x$ is decreased
below a critical value $x_c$, the Fermi surface
is rearranged (Fig \ref{fig:Fermi surface}),
leading to changes in $R_H$. These changes were discussed,
in the context of a theory of $Cr$,
by Bazaliy {\it et al}, \cite{Bazaliy04}
and the related discussions, in the context of
the d-density wave model of cuprates,
were given by Ref. \onlinecite{Chakravarty02}.
The new results presented here include consideration
of a wide range of dopings and gap values
(Fig \ref{fig:hall}),
a systematic treatment of all relevant effects
(Sec. \ref{sec:Formalism}),
a discussion of the weak field-strong field crossover
(Sec. \ref{sec:Weak Field Limit}),
a treatment of precursor effects
(Sec. \ref{sec:Normal State Region}),
and detailed application of the model to the
electron-doped cuprates
(Fig \ref{fig:hall} and conclusion).
We note, unfortunately,
that the specifics of the critical behavior
and weak to strong field crossover will be visible
(at presently accessible fields $B\lesssim 50$T)
only at samples doped very close ($\delta x\sim 0.005$)
of the critical doping,
and at the present level of sample purity,
these effects will be complicated by magnetic breakdown.
The data of Ref. \onlinecite{Dagan04}
suggest $x_c\approx0.165$
while the data of Ref. \onlinecite{Onose01}
indicate $x_1>0.1$ and is most probably $\sim 0.12$.
Thus the gap apparently opens quite fast with doping.
However, important discrepancies remain
between data and calculation.
At $x<x_1$ our calculation shows $R_H\approx 1/x$
(as expected for a Fermi surface with one electron pocket
containing $x$ electrons).
The data of Ref. \onlinecite{Onose01} are qualitatively
consistent with this trend, but are offset,
suggesting that the actual doping is slightly lower
than the nominal one.
The $x=0.11$ data point of Ref. \onlinecite{Dagan04}
is qualitatively inconsistent with theory.
Also, the magnitude reported \cite{Dagan04}
of $R_H$ at $x>x_c$ is inconsistent with theory.
One expects that as $x$ is increased beyond $x_c$,
$R_H$ should revert to the band value, $R_H\sim 1/(1-x)$.
While the sign and order of magnitude are correct,
the numerical value is too large;
there is no reasonable choice of band parameters
which reproduces the magnitude to better than a factor of two.

Our work suggests several important directions
for future research. On the experimental side,
we suggest that our finding that $R_H\approx 1/x$
for $x<x_1$ can be used
to determine the actual doping of lightly doped samples
($x\lesssim 0.1$).
We also note that experiments which pinned down the behavior
in the $x_c>x>x_1$ regime and more precisely located $x_1$
would be very helpful.
Finally, clarifying the behavior at $x>x_c$ is
potentially of considerable interest,
as it gives (in principle) evidence of the modification
of the Fermi suface by finite range SDW correlations.
On the theoretical side,
important directions include combining the present
considerations with those of Ref. \onlinecite{Oganesyan04}
to obtain a theory of the Nernst effect,
and improving the theoretical treatment of the SDW precursor
effects. Work on the AC Hall effect is in progress.


{\it Acknowledgements}
We thank R. Greene, Y. Dagan and M. R. Norman
for helpful discussions and A. J. Schofield
for drawing our attention to the strong-field crossover
and the possibility of magnetic breakdown.
This work was supported by NSF-DMR-0431350.

\end{document}